\documentclass[conference]{IEEEtran}
\IEEEoverridecommandlockouts

\usepackage{cite}
\usepackage{amsmath,amssymb,amsfonts}
\usepackage{algorithmic}
\usepackage{graphicx}
\usepackage{textcomp}
\usepackage{xcolor}
\usepackage{balance}
\usepackage[most]{tcolorbox}
\def\BibTeX{{\rm B\kern-.05em{\sc i\kern-.025em b}\kern-.08em
    T\kern-.1667em\lower.7ex\hbox{E}\kern-.125emX}}
\tcbset{textmarker/.style={
        enhanced,
        parbox=false,boxrule=0mm,boxsep=0mm,arc=0mm,
        outer arc=0mm,left=5mm,right=3mm,top=7pt,bottom=7pt,
        toptitle=1mm,bottomtitle=1mm}}
\newtcolorbox{noteBox}{textmarker,
    borderline west={4pt}{0pt}{gray},
    colback=gray!10!white}
\newcommand{\note}[1]{\begin{noteBox} #1 \end{noteBox}}

\begin{document}

\title{Task-Based Role-Playing VR Game for Supporting Intellectual Disability Therapies}

\author{\IEEEauthorblockN{WenChun Chen}
\IEEEauthorblockA{
\textit{Technical University of Munich}\\
Heilbronn, Germany \\
wenchun.chen@tum.de}
\and
\IEEEauthorblockN{Santiago Berrezueta-Guzman}
\IEEEauthorblockA{
\textit{Technical University of Munich}\\
Heilbronn, Germany \\
s.berrezueta@tum.de}
\and
\IEEEauthorblockN{Stefan Wagner}
\IEEEauthorblockA{
\textit{Technical University of Munich}\\
Heilbronn, Germany \\
stefan.wagner@tum.de}
}

\maketitle

\begin{abstract}
Intellectual Disability (ID) is characterized by deficits in intellectual functioning and adaptive behavior, necessitating customized therapeutic interventions to improve daily life skills. This paper presents the development and evaluation of Space Exodus, a task-based role-playing Virtual Reality (VR) game designed to support therapy for children with ID. The game integrates everyday life scenarios into an immersive environment to enhance skill acquisition and transfer. Functional tests and preliminary experiments demonstrated the system's stability, usability, and adaptability, with 70--80\% of participants demonstrating successful skill transfer to new challenges. 

Challenges, such as VR discomfort, controller misoperation, and task complexity, were identified, emphasizing the need for ergonomic improvements and adaptive guidance. The results provide empirical evidence supporting VR as a promising tool in ID therapy. Future work will focus on refining gameplay mechanics, enhancing user guidance, and expanding accessibility to broader populations.
\end{abstract}

\begin{IEEEkeywords}
Virtual reality, intellectual disability, therapeutic interventions, transfer learning, task-based role-playing games.
\end{IEEEkeywords}

\section{Introduction}

Intellectual Disability (ID) is a condition marked by deficits in intellectual functioning and adaptive behavior \cite{american2000diagnostic, tasse2021american}. The global prevalence of ID is estimated to be between 1--2\%, and it is usually diagnosed when a child exhibits significant developmental delays compared to their peers \cite{world2023global}. Intellectual functioning encompasses the capacity to learn, think, and understand, while adaptive behavior is divided into conceptual, social, and practical skills. Given that ID is not a curable condition, assistance strategies focus on skill acquisition to enhance the individual's ability to live as independently as possible. Therefore, one key aspect of rehabilitation is teaching self-care skills \cite{katz2008intellectual}.

Virtual Reality (VR) is a simulated experience that immerses users in a three-dimensional environment, allowing them to interact with a virtual world without being physically present \cite{vince2011introduction}. Studies have shown that VR holds great potential as a supportive tool for children with ID \cite{bailey2022virtual, standen2005virtual}. One reason is that children with ID often require multiple scenarios and additional practice to grasp new concepts due to memory deficiencies affecting short- and long-term retention \cite{vicari2016memory}. Because these children are vulnerable to information overload, VR scenarios should be designed to include only essential objects relevant to the learning task and provide straightforward instructions \cite{giuliani2015vision}. 

Moreover, VR offers unique advantages over real-world teaching environments, such as unlimited patience, which can positively impact the self-esteem of users \cite{weiss2003virtual}. VR's highly customizable and low-complexity nature allows the simulation of various situations, making it an effective tool in the occupational therapies of children with ID \cite{ahn2021combined}. However, despite these benefits, users may experience visual discomfort, VR sickness, anxiety, or claustrophobia during use, which can lead to reduced user engagement, increased dropout rates, or limited session durations, thereby impacting the overall effectiveness of VR interventions \cite{simon2024cybersickness}. Nonetheless, careful design can make VR a valuable asset supporting ID therapy.

This paper presents the development of a task-based role-playing VR game designed to conduct experiments that provide empirical evidence on the efficacy of VR in the therapy of children with ID. Given the vulnerable nature of the target population, ethical considerations have been a cornerstone of this study.

This paper begins with an introduction section that provides background information on ID, outlining the challenges associated with rehabilitating children with this condition and proposing VR as a potential solution. The Related Work section reviews relevant academic literature and perspectives on using this technology in therapeutic contexts for children with ID. The methodology section details the requirement elicitation process and system design, outlining how the VR game was tailored to meet the specific needs of children with ID and the experiment, tracking its progression, and providing insights into the implementation. Finally, the paper concludes by presenting the experiment's results, summarizing key findings, and discussing future work that could further enhance the use of VR in ID rehabilitation.

\section{Related work}\label{RW}

VR has great potential to help individuals with ID. Simulating different scenarios allows users to practice and learn skills in a safe and engaging environment, making it a powerful tool for teaching and training \cite{torrey2010transfer, michalski2023improving, barnett2002and}. A few studies that have developed games and evaluated the effectiveness of transfer learning highlight the promising potential of this approach.

The study by Vasconcelos et al.~\cite{vasconcelos2017protocol} focused on using VR to improve literacy rates among students with ID. They developed a game called ``Aprendendo com Trafas,'' which categorizes everyday objects based on their purpose and integrates them into daily activities to help children learn through visual memory. These activities include shopping for items for a picnic, family trip, or surprise party at a mall. The game showed positive results in creating an interactive learning environment through VR, but it faced challenges due to a limited vocabulary and the need for scalability. 

The study by Cherix et al.~\cite{cherix2020training} conducted transfer learning to simulate various learning scenarios related to road crossing, including judgment of when and where to cross and different variables like weather, day/night time, and driver behavior. The experiment involved 15 individuals with ID aged 9--18. Results showed that the VR simulator was effective only for four consecutive sessions across sixteen crossings. This demonstrated the potential of VR in transfer learning applications. However, the transfer ability from the virtual world to the real world has yet to be measured, requiring further experimentation.

The research by Michalski et al.~\cite{michalski2023improving} evaluated the possibility of transfer learning in waste management using VR for individuals with ID. Participants were asked to sort waste based on its characteristics into specific trash cans, such as general waste, recycling, and garden and food organics. The experiment had three phases: a pre-test and a post-test in the real world, followed by a learning phase with VR tutorials. In Stage 1, participants sorted objects by color, and then in Stage 2, they were sorted into the appropriate trash cans in the VR environment. The results showed that 19 out of 32 participants met the learning targets, encouraging further studies with larger sample sizes and additional experiments to solidify the effectiveness.

The effectiveness of transfer learning using VR for individuals with ID in rehabilitation has shown promise, but its potential far exceeds current implementations \cite{haryana2022virtual}. Therefore, further research and practical applications are needed to solidify these findings. Additionally, past approaches have primarily focused on educational support, which can be monotonous and unsustainable in the long run. This research proposes a task-based role-playing game that incorporates everyday life scenarios into tasks, aiming to balance playfulness with supportiveness in rehabilitation.

\section{Methodology}\label{M}

\begin{figure}[htbp]
\centerline{\includegraphics[width=9cm]{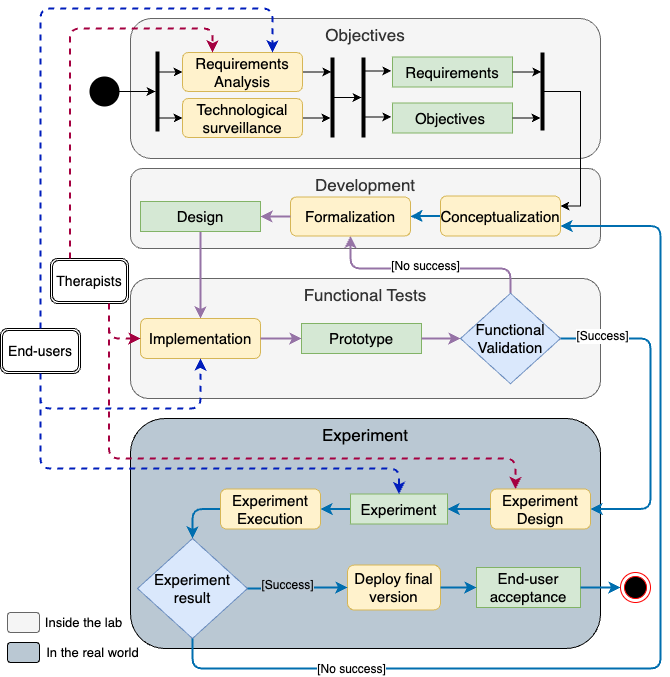}}
\caption{Activity diagram illustrating the user-centered research process \cite{berrezueta2022user}.}
\label{fig:activitydiagram}
\end{figure}

This research aims to develop and evaluate a task-based role-playing VR game as a therapeutic tool for supporting children with ID by enhancing their adaptive and cognitive skills through immersive and interactive learning experiences. The activity diagram in Figure \ref{fig:activitydiagram} illustrates the user-centered research process of this study. 

In the objectives phase, the requirement analysis is conducted through a literature review and expert interviews to evaluate current academic research and clinical practices. Based on this analysis, we define the functional, non-functional, and pseudo requirements\footnote{These requirements usually do not impact the user experience or software's functionality but can influence its design, deployment, or operation \cite{wiegers2013software}.} and set clear objectives for the VR game. 

In the development phase, these ideas are conceptualized and used to design the system. Next, the system is implemented during the functional test phase, and the game undergoes testing to ensure it meets the defined requirements. Finally, we design and conduct a preliminary experiment to evaluate the system, followed by deploying the game's final version.

\subsection{Requirements Elicitation}

The requirement elicitation phase included interviews, surveys, collaborative brainstorming, prototyping, and observational methods \cite{tiwari2012selecting}, involving 15 experts: five occupational therapists, five clinical psychologists, and five special education teachers. Each expert, with extensive experience in therapy for children with ID, participated in 45- to 60-minute sessions focused on defining key therapeutic objectives, customization needs, and practical instructional design practices for VR. The experts also provided valuable insights into communication techniques tailored to individuals with ID, including specific methods for giving instructions. This diverse group of participants ensured an inclusive approach to designing therapeutic goals and customization features for the VR system.

These sessions reveal that therapists find VR not widely used in daycare centers or physical therapy for two reasons. First, VR technology is considered too expensive, and the supporting evidence for its effectiveness is currently insufficient. Second, rehabilitation for children with ID requires high levels of customization due to the variability in adaptive function deficiencies. At the end of the requirement elicitation, we identified 21 Functional Requirements, 15 Non-Functional Requirements, and 7 Pseudo Requirements. Below, we present the list of these requirements more broadly. 

\textit{Functional Requirements}
\begin{itemize}
  \item The game is a role-playing game. Players are tasked with escaping a spaceship in this immersive space survival game.
  \item The game should have five tasks. Each task shall require the previously learned skill and some new skills.
  \item The difficulty of the task shall gradually increase.
  \item The game shall have a main menu and a pause menu. In the main menu, players can customize their preferences. The pause menu allows players to pause the game whenever needed.
\end{itemize}

\textit{Non-Functional Requirements}
\begin{itemize}
  \item Our game design incorporates numerous hand-eye coordination exercises, as research has shown that such physical activities can improve attention span in children \cite{MONNO2002187}.
  \item By leveraging VR's ability to simulate various virtual scenarios, the game should be designed to combine with daily life tasks.
  \item Presenting instructions clearly and concisely is essential. 
  \item The design of the narratives should be patient and allow ample response time after giving instructions.
\end{itemize}

\textit{Pseudo Requirements}
\begin{itemize}
  \item Kids using VR should be at least 8 years old to understand the game and use the controllers.
  \item The game can be played on Android VR devices.
  \item The biggest concern with VR usage is its adverse effects, particularly VR sickness. Taking a break every 20 minutes is recommended; gradually increasing playing time is a preventative measure.
  \item It is suggested to prepare an obstacle-free zone when using VR, as users cannot see their surroundings while wearing VR devices.
\end{itemize}

 We developed our research hypothesis by combining insights from academic research and clinical practice taken from the requirement elicitation. \textit{A task-based role-playing game focused on daily life skills may enhance playfulness while maintaining the effectiveness of transfer learning, compared to a more educational lean game}.

\subsection{Materials}
The VR device used during development is the Meta Quest 2 headset, known for its portability, high-resolution display, and wireless design. It offers 1832 x 1920 pixels per eye, a 90Hz refresh rate, and is powered by a Snapdragon XR2 processor. Built-in tracking and ergonomic controllers provide a seamless, immersive experience without external sensors, making it ideal for various applications, including gaming and therapy \cite{meta2022}. The VR game is developed using the Unity platform, which offers flexibility and robust support for VR applications. At the same time, the programming is done in C\#, a versatile language that enables efficient development of interactive and dynamic elements within the VR environment \cite{unity2024}. This combination allows for scalable and customizable experiences tailored to the specific needs of children with ID. 

\subsection{Description of the game}

\begin{figure}[htbp]
\centerline{\includegraphics[width=8cm]{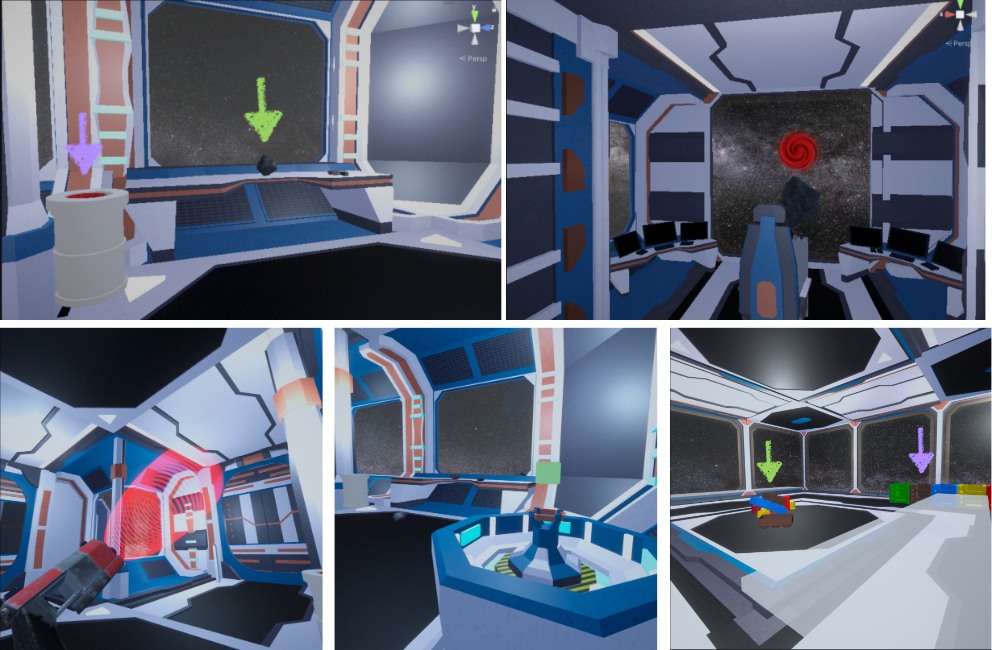}}
\caption{Screenshot of the \textit{Space Exodus} scenarios where the tasks are performed.}
\label{Captures}
\end{figure}

The game's name is \textit{Space Exodus}.\footnote{GitHub repository: https://github.com/TUM-HN/Space-Exodus} The game tasks are designed to mirror real-world tasks, helping players with ID practice essential life skills in a space-themed virtual environment, as shown in Figure \ref{Captures}. 

\textit{Task 1 -- Throw the Meteor into the Trashcan.}
Simulates basic waste disposal and motor coordination by grabbing and moving objects to a designated location. This activity reinforces hand-eye coordination and the concept of sorting waste, which is essential for daily and environmental awareness.

\textit{Task 2 -- Clean up the Mess in the Bricks Room.}
Mirrors tidy up and organize skills by requiring players to sort items into specific bins. This task enhances attention to detail, categorization, and repetitive action management, like organizing and cleaning a living or classroom space.

\textit{Task 3 -- Disintegrate the Large Meteor.}
It represents breaking down large objects or recycling items using a "tool" (the gun) to disintegrate the meteor, a valuable skill in real-world activities like breaking down boxes or managing to recycle.

\textit{Task 4 -- Place Energy Source on Analyzer.}
It involves identifying and handling objects with a specific purpose, akin to sorting and categorizing items for tasks like cooking, organizing school supplies, or assembling items. It also promotes spatial awareness and the recognition of tools by function.

\textit{Task 5 -- Pilot the Spaceship to the Wormhole.}
This task mimics navigation and decision-making under pressure, requiring attention, quick reflexes, and planning to avoid obstacles. These skills have real-world applications in tasks like crossing streets, navigating crowded spaces, and improving situational awareness, making it valuable for adaptive functioning in children with ID.

The game enhances practical skills and provides an engaging and meaningful therapeutic experience by associating these in-game tasks with everyday activities. This approach helps players develop abilities they can apply outside the virtual environment, making the game a valuable tool for skill development.

\subsection{Functional Test}

 Demographic data were collected to ensure the test's applicability across diverse user profiles. This functional test was conducted and supervised by the experts who also participated in the requirement elicitation phase. The test covered every game aspect, including the menu, interactions, timeline, and customization features. Each button in the main menu was verified for functionality, the locomotion system, user interface, and audio and haptic feedback were assessed for interactions, the timeline system was tested for smooth progression, and the customization section was reviewed for language options, control methods, and subtitle features. This thorough testing process ensures the reliability and effectiveness of the VR game. This functional test focuses on: 

\begin{itemize}
    \item \textit{Basic functionality} to verify the accuracy of game controls, navigation, and interactions within the virtual environment.
    \item \textit{Performance and stability} to verify the smooth gameplay and system stability without crashes. 
\end{itemize}

\subsection{Experiment}

\begin{figure}[htbp]
\centerline{\includegraphics[width=8.5cm]{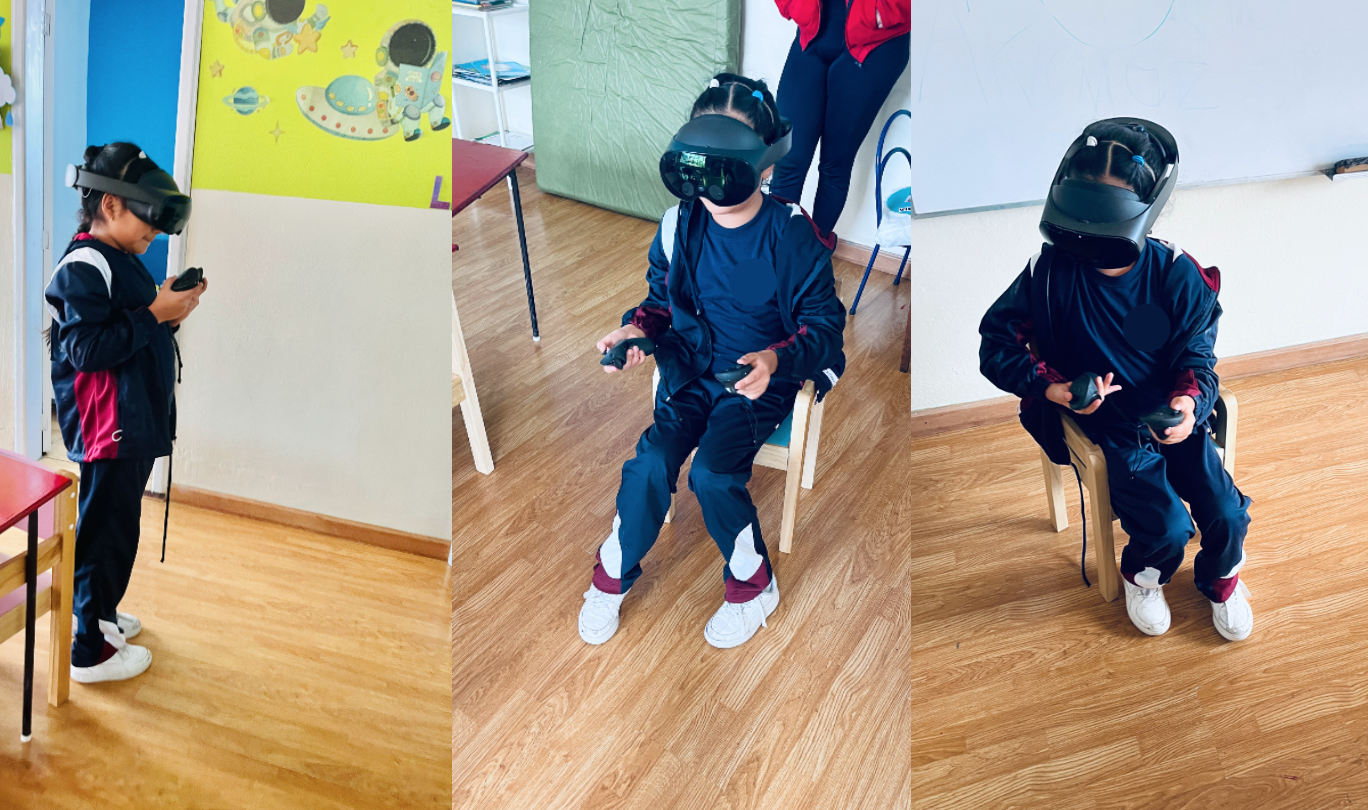}}
\caption{Children performing a functional test of the VR game, interacting with virtual objects using hand-tracking technology. Conducted under ethic-approved guidelines.}
\label{Test}
\end{figure}

The experiment involved eight children aged 8–12 years, all diagnosed with varied severity of ID. Each session was conducted in a controlled indoor environment with a designated play space of 1.5 square meters (see Figure \ref{Test}). Participants could play standing or seated, with movement and rotation managed through controller manipulation to accommodate diverse physical needs.

Therapists supervised the sessions, observing the children’s interactions with the game to collect qualitative data on task performance, emotional responses, and overall engagement. Developers also gathered feedback to refine the game based on real-world usage. Comprehensive guidance was provided during the initial playthrough to familiarize participants with the mechanics and objectives of the game. This experiment focuses on:

\begin{itemize}
    \item \textit{Adaptive learning and customization} to ensure the game adjusted difficulty based on user progress and allowed personalization according to therapeutic needs. 
    \item \textit{Cognitive and emotional engagement} to assess the clarity of instructions and emotional feedback, ensuring appropriate reinforcement. 
\end{itemize}

Participants worked through all five tasks, with data on completion times and observed behaviors informing adjustments to task design. These insights will guide future iterations, with plans to incorporate objective metrics such as response accuracy and system intuitiveness. This pilot study is the foundation for future experiments with larger participant groups, including collaborations with the UNESCO Chair in Ecuador to expand the study’s scope and diversity.

\subsection{Ethics Compliance}
This study was conducted under the approval of the Institutional Review Board (IRB) to ensure the ethical treatment of all participants, particularly considering the vulnerable nature of ID children. All protocols adhered to international ethical standards for research involving human participants, prioritizing safety, informed consent, and inclusivity. Future studies will continue to build upon these ethical foundations, emphasizing the responsible development and implementation of VR-based therapeutic tools.

\section{Results}\label{R}
This preliminary study evaluates a role-playing VR game's effectiveness in improving the concentration of children with ID. The game, designed as an assistive tool, aims to enhance concentration, which we hypothesize will benefit their daily living skills. While final results are pending, we will present initial findings from functional tests followed by preliminary observations of the experiment.

The results of the functional tests and the experiment confirmed the overall effectiveness and stability of the VR system. All controls and navigation operated smoothly, allowing users to interact with the virtual environment without issues. During the experiment, the system showcased excellent stability, with only minor issues observed that did not detract from the overall user experience. Minor concerns emerged: occasional controller input delays during precise tasks and infrequent unresponsiveness in menu navigation. Despite these isolated occurrences, the system performed reliably, with no significant crashes, rendering problems, or synchronization errors, reinforcing its robustness and suitability for therapeutic applications. These findings highlight the game’s strong foundation and readiness for broader implementation.

In the experiment phase, the evaluation of the VR game in RPG format yielded several positive outcomes, demonstrating its potential to engage children while supporting skill development. Most participants could apply skills learned in earlier tasks to later ones, with 70–80\% successfully transferring knowledge, such as object manipulation from Task 1, to new challenges in Task 2. The immersive and dynamic nature of the game maintained engagement, with most children expressing a willingness to continue playing. Task 5 (Pilot the spaceship) was particularly popular, thanks to its interactive and rewarding design. 

On average, completing all tasks during the first playthrough took 20–35 minutes, with well-calibrated difficulty adjustments ensuring that progression aligned with the players' proficiency levels. Comprehensive guidance was provided throughout the initial playthrough to support task completion and familiarize children with the game mechanics. In subsequent playthroughs, the guidance was gradually reduced to encourage independent problem-solving, and the time required to complete each task varied depending on its complexity, reflecting the adaptive design of the gameplay. 

Task 1 (Throw the Meteor) was the quickest (2–4 minutes). Task 4 (Place energy source) was the most time-consuming (4–8 minutes) due to its route planning and precision requirements, e.g., the ladder climbing, in accessing the room where Task 5 is located. While the game encouraged focused problem-solving, 70\% of participants prioritized task completion over exploration, indicating that the game effectively guided attention toward its objectives. 

VR discomfort, claustrophobia, and the weight of the helmet were considered the main reasons why children with ID were reluctant to use the game. However, the well-lit, colorful, and open spaces with calming background music created a low-anxiety environment, counteracting these concerns. 

Despite these strengths, some challenges emerged, offering opportunities for improvement. Some children struggled with a lack of direct instructions and complex hand-eye coordination tasks, leading to occasional frustration or disengagement after 7–10 minutes of repeated failure (approximately 3-4 attempts). Navigating the spaceship environment proved challenging for some, with players taking 2–5 minutes to locate task areas due to a lack of intuitive visual or auditory cues.

\section{Discussion}\label{D}

The findings of this study demonstrate the potential of task-based role-playing VR games, such as Space Exodus, to serve as practical therapeutic tools for children with ID. 

\note{\textbf{Finding 1 -- Engagement through gamification:} The dynamic and task-based structure successfully combined therapeutic goals with engaging gameplay, maintaining user interest and motivation.}

The game successfully engaged participants, supporting the development of cognitive and motor skills while fostering skill transfer across tasks. The results align with previous research emphasizing the utility of immersive technologies in therapeutic contexts \cite{bailey2022virtual, standen2005virtual}, suggesting that task-based role-playing VR-based interventions can enhance traditional therapy methods.

\note{\textbf{Finding 2 -- Need for improved hardware usability:} VR discomfort and controller misoperation highlight the importance of ergonomic hardware and intuitive interfaces.}

\pagebreak

The game demonstrated strong user engagement and adaptability, but the identified challenges highlight the need for usability enhancements aligning with findings in \cite{simon2024cybersickness, buhler2018reducing}. The prevalence of VR-related discomfort and difficulties with controller operation emphasizes the importance of using ergonomic, user-centered hardware and more intuitive interfaces.

\note{\textbf{Finding 3 -- Introduce reward systems to mitigate frustration: } Repetitive tasks are necessary for practicing previously learned skills, but these often lead to boredom.}

While reducing repetitions is not recommended, incorporating a reward system after each successful move can increase players’ willingness to practice \cite{schultz2007reward}, \cite{hidi2016revisiting}. This approach helps sustain motivation and mitigates the negative perception of repetitiveness. Additionally, progressive hint systems are crucial to provide support during failures, as results showed that disengagement often occurred after 3–4 consecutive failed attempts, leading players to abandon the game.

\note{\textbf{Finding 4 -- Enhance adaptive task guidance and cues} to improve task completion and reduce frustration.}

Specific tasks indicate additional support to sustain engagement. Task 4 (Place energy source + Ladder climbing) required the most time to complete due to route planning. In contrast, Task 5 (Pilot the spaceship) benefited from hints to aid understanding and navigation. These findings suggest a need for adaptive guidance, explicit instructions, and streamlined navigation to maintain engagement and reduce frustration, highlighting the game's potential for children with ID.

\note{\textbf{Finding 5 --  Increase Task Variety:} to enhance engagement, especially after players have completed all tasks.}

Children with ID preferred high-variety tasks like Task 5: Pilot Spaceship for its dynamic challenges, while repetitive tasks like Task 2: Clean Up the Mess were less appealing when revisited. To keep players motivated, incorporating a greater diversity of tasks, integrating challenging levels as standalone reward stages, and adding a reward system for successful performance are recommended to sustain interest and provide a sense of accomplishment. Additionally, designing challenging tasks as standalone reward levels after completing all stages could effectively maintain interest and provide a sense of accomplishment aligning with the finding in \cite{kammermann2024virtual}.

\pagebreak

\section{Conclusions}\label{C}

This study demonstrates the potential of task-based role-playing VR games as practical therapeutic tools for children with ID. By incorporating everyday life scenarios into an immersive virtual environment, the developed game, \textit{Space Exodus}, effectively engaged participants while fostering the transfer of learned skills to new challenges. The results of the experiments confirmed the system’s stability, usability, and adaptability, with most children successfully applying previously acquired skills to later tasks. The game's dynamic nature, particularly Task 5 (Pilot the spaceship), proved highly engaging and enjoyable for participants, underscoring the value of interactive and rewarding design.

While the results highlight the game’s ability to sustain attention and promote cognitive and motor skill development, some challenges were identified—barriers to VR use, such as headset discomfort, controller misoperation, and limited user engagement. In gameplay, difficulties with complex instructions, hand-eye coordination, and task navigation occasionally led to frustration or disengagement, particularly during more demanding tasks. These findings emphasize the need for ergonomic design, introduce reward systems, enhance adaptive task guidance, and increase task variety to enhance the user experience and reduce frustration. Addressing these issues will not only improve accessibility but also ensure that the system effectively supports the diverse needs of children with ID.

This study provides robust empirical evidence supporting the use of VR in ID therapy, highlighting its potential to enhance skill development and engagement. Future efforts will focus on integrating adaptive feedback systems, refining user interfaces, and diversifying tasks to address the challenges identified. Long-term studies with more extensive and diverse populations are needed to validate these findings and explore the broader applications of VR in therapeutic contexts. Additionally, future research should prioritize cost-effective implementation strategies, such as deploying games on affordable VR platforms or integrating VR therapy into educational and healthcare programs. This work represents a significant step toward leveraging immersive technology to improve the lives of children with ID.

\balance
\bibliographystyle{ieeetr}
\bibliography{Paper_AIxVR}

\end{document}